Prolégomènes à une post-esthétique des imaginations artificielles
*(Pour une critique phénoménologique
de la pratique esthétique liée aux IA)*

Philippe Boisnard
Laboratoire Paragraphe - Paris VIII

Résumé : L'accélération de l'usage des intelligences artificielles (IA) génératives, depuis 2015 et le tournant opéré par *Deepdream*, tend à occulter une véritable analyse de ce que l'on pourrait définir comme *imagination artificielle*. Les IA étant, soit réduites à de simples instruments, soit pensées selon une forme de techno-théologisme. Notre recherche tend à suspendre toute forme de jugement pour saisir phénoménalement l'émergence de ces IA. En reprenant la question de l'esthétique de Hegel et de l'art comme libre production de l'esprit, mais en la déplaçant vers la question des IA génératives et dès lors d'une post-esthétique, cet article va montrer la spécificité phénoménale des images générées par IA.

Summary: The acceleration to use of generative artificial intelligence since 2015 and the turning point made by Deepdream, tends to obscure a real analysis of what we could define as artificial imagination. AI are either reduced to simple instruments, or thought of according to a form of techno-theologism. Our research tends to suspend any form of judgment to phenomenally grasp the emergence of these AIs. By taking up the question of Hegel's aesthetics and art as the free production of the mind, but by moving it towards the question of generative AI and therefore a post-aesthetics, this article will show the phenomenal specificity AI-generated images.

*« Peut-on parler de la technique sans la mythifier (soit favorablement, soit défavorablement) ? Peut-on parvenir à analyser lucidement ses réalités et ses possibilités, en appréciant la marge de mythification qu'elle véhicule inévitablement ? Comment reposer le problème des limites de la technique en évitant à la fois le ton « grand seigneur » d'une philosophie prétendument souveraine et le mimétisme techniciste d'un discours néo-positiviste qui ne ferait que célébrer les exploits techniques sans penser ni leur condition de possibilité ni leur portée ? »*
Dominique Janicaud, Les limites de la technique, mythes et réalités, noesis 29, 2017.

*« Nous ne sommes plus les seuls à voir et à générer les images. »* Laurence
Danguy, Julien Shuh, L'oeil Numérique : vers une culture visuelle hybride, Sociétés et représentations, éditions de la sorbonne, 2023.

Les discours accompagnant l'émergence des IA génératives depuis 2019 sont nombreux, portant très souvent sur la question de la concurrence homme-algorithme, discours tantôt fascinés, tantôt inquiets, suspicieux quant aux horizons du développement accéléré de l'usage des IA[1].

Plutôt que d'entrer dans un tel débat, il serait plus pertinent de circonscrire préalablement la nature des objets dont nous parlons et donc d'établir la possibilité d'une approche qui pourrait les définir.

La question de la comparaison/compétition entre production humaine et production algorithmique provient du fait mimétique qui prédéfinit, semble-t-il, les productions générées des IA telle Midjourney, Dall-e 2, Stable diffusion, Photoshop pour la génération d'images, ou bien chat-GPT pour la génération de textes.

En effet, et cela a été le moteur de leur expansion et de leur autorité : si les premières IA génératives depuis *Deepdream* n'allaient pas forcément vers un mimétisme graphique, mais bien plus vers des formes combinatoires que l'on a désigné de surréaliste[2] ou de psychédélique[3], peu à peu ce qui prévaut, c'est la possibilité de reproduire, non pas le réel, mais la logique et l'esthétique du réel qui domine la représentation humaine.

La question concurrentielle provient du fait que l'on rabat apriori la production générée sur la production faite par l'homme. Il ne faut pas nier que ces questions se posent, car elles interrogent la substitution pour certaines tâches de l'humain par des IA, notamment au niveau de la production en masse de textes ou bien d'images[4].

---

[1] Nous pourrions résumer cela en un dualisme opposant un réductionnisme transcendantal de l'IA et une forme de transcendance techno-théologique. Le réductionnisme, par son champ lexical (celui du Perroquet : Garry Marcus, Hervé Fischer), renvoie à la compréhension cartésienne de la machine dans *Le Discours de la méthode*. « Mais Siri ne pense pas. C'est la condition fondamentale de son bon fonctionnement. Il n'est qu'un programme disposant d'un nombre grandissant de recettes algorithmiques. Mon perroquet pense certainement plus que Siri, choisissant le moment de parler ou de se taire et la séquence sonore mémorisée qu'il veut, même lorsque je ne lui demande rien. Mon perroquet a plus d'intelligence que le plus puissant des ordinateurs qu'on inventera ja- mais. Pourquoi ? : mon perroquet est un sujet, tandis qu'un ordinateur ne sera jamais qu'un objet, même intégré dans un robot imi- tant à la perfection les formes d'un humanoïde. » Hervé Fischer, *Mythanalyse de l'intelligence artificielle*.
« Les pies et les perroquets peuvent proférer des paroles ainsi que nous, et toutefois ne peuvent parler ainsi que nous, c'est-à-dire en témoignant qu'ils pensent ce qu'ils disent. (…)
On peut bien concevoir qu'une machine profère des paroles; mais non pas qu'elle les arrange diversement pour répondre au sens de tout ce qui se dira en sa présence, ainsi que les hommes les plus hébétés peuvent faire» Descartes, *Discours de la méthode*.
La projection techno-théologique est pour sa part développée selon une intention transhumaniste (Jean-Marc Moschetta, Anthony Levandowski).

[2] https://www.philomag.com/articles/lintelligence-artificielle-reve-t-elle-de-poesie Cet article du 14 Janvier 2021, réfléchit sur la la relation entre le surréalisme et la production d'IA générative dans le processus de Bots of New York. https://www.facebook.com/botsofnewyork/. Voir aussi : Le texte inaugural de Jill Gasparina, Le surréalisme et la peinture (reboot) dans Peinture : obsolescence déprogrammée, Licences Libres, 2022, hal-03606345 : « une coïncidence formelle, à savoir les similitudes visuelles entre les œuvres de certains peintres surréalistes et les images produites aujourd'hui par les artistes mettant en jeu des intelligences artiffcielles. »

[3] Ainsi, David Auerbach écrit-il dès le mois de juillet 2015 : « Le mois dernier, Google a introduit un programme trippant appelé DeepDream, que les premiers utilisateurs ont utilisé pour générer des variations psychédéliques de tout, de Hieronymus Bosch au porno. ». https://slate.com/technology/2015/07/google-deepdream-its-dazzling-creepy-and-tells-us-a-lot-about-the-future-of-a-i.html

Toutefois, ce cercle qui repose sur la comparaison ne permet pas de réfléchir spécifiquement la nature de la génération automatique par IA. Trop souvent, seul le résultat est perçu et non plus le processus ou bien les composantes du processus. S'il s'agit comme l'exprime Jacques Taminiaux dans la citation d'exergue de penser les possibilités propres de la technique, il faut l'abstraire tout à la fois du réductionnisme instrumental et de la fascination anthropo-transcendante. En quelque sorte, il s'agit de l'émergence des IA en tant que *machine célibataire* tel qu'elle est définie par Deleuze et Guattari : pour une machine célibataire, « l'essentiel est l'établissement d'une surface enchantée d'inscription ou d'enregistrement qui s'attribue toutes les forces productives et les organes de production, et qui agit comme quasi-cause en leur communiquant le mouvement apparent »[5].

Ce que nous proposons alors c'est de réfléchir, en un sens phénoménologique, au déplacement de la question esthétique hegelienne vers une possibilité post-esthétique. Si *l'esthétique* s'est constituée comme une approche phénoménologique de la libre création des oeuvres de l'esprit humain, en quel sens définir une *post-esthétique* amènerait à poser une approche philosophique des productions issues des IA comme distinctes de celles des agents humains et devant ouvrir alors à une phénoménologie spécifique de ses processus.

Pour ce faire, notre approche s'établit sur un déplacement du terme d'IA en tant qu'intelligence artificielle, vers l'imagination artificielle. Le terme d'imagination artificielle n'est pas nouveau, il est introduit au niveau de la formation par des techniques d'image par A. Kaufman[6] pour signifier : « l'élaboration d'assemblages, plus ou moins complexes, par un programme réalisant une exploration (déterministe ou aléatoire) d'un univers combinatoire, contenant un nombre généralement très élevé de tels assemblages ». Puis il est lié à la possibilité de machines imaginatives en 2001 par Igor Aleksander[7], pour ensuite apparaître strictement relié aux IA à partir de 2015 et de l'avènement de *deepdream* par exemple dans l'article de Joaquim Sivestre, Yasushi Ikeda et François Guéna qui interroge les processus d'une

---

[4] Cette question du remplacement si elle a émergé vers 2017-2019 (https://www.blogdumoderateur.com/place-intelligence-artificielle-design/), est devenue omniprésente surtout à partir de 2023 du fait de l'accélération des évolutions des IA et de leur interfaçage grand public. Cf. https://www.bfmtv.com/tech/les-intelligences-artificielles-vont-elles-remplacer-les-artistes_GN-202301120101.html article de janvier 2023. Cette question de la substitution se pose d'autant plus lorsqu'il s'agit de la substitution d'un créateur vivant par un usage mimétique d'IA. C'est par exemple ce qui se produit pour Greg Rutkowski, comme l'explique Melissa Heikkilä dans le MIT Technology Review en 2022, https://www.technologyreview.com/2022/09/16/1059598/this-artist-is-dominating-ai-generated-art-and-hes-not-happy-about-it/ .

[5] G. Deleuze, F. FGuattari, Anti-Œdipe, ed. Minuit, 1972. Cf. Philippe Boisnard, La machine célibataire et l'IA, 2022, http://databaz.org/xtrm-art/?p=862

[6] A. Kaufman Publié dans la Revue française d'informatique et de recherche opérationnelle, t. 3, n° 3, 1969, p. 5-24.

[7] Igor Aleksander, *How to build a Mind, Towards machine with imagination*, Colombia University Press, 2001.

« imagination artificielle » pouvant générer des architectures à partir de ConvNet[8]. À partir de 2017, Grégory Chatonsky, de même, réfléchit spécifiquement l'imagination artificielle en liaison aux IA, notamment et surtout aux IA génératives d'images, en posant la question de cette imagination selon une nécessité relationnelle : la liaison entre un apprentissage machiniste/automatisé fondé sur la mémoire humaine (celle des nos images, textes déposés numériquement)[9] et la relation que nous avons avec les images alors générés/hallucinées par les IA.

Si pour une part nous nous situons dans de tels horizons, au sens où le terme d'artificialité renvoie à la question synthétique de la production d'images d'images, toutefois, notre questionnement, en suspendant en quelque sorte la relation à l'homme, tente de saisir la phénoménalité des images générées par les IA.

Ce déplacement phénoménologique aboutira ainsi à comprendre entre autre, de quel espace s'agit-il quant à l'espace représenté par une IA ? Quelle nature temporelle est à l'oeuvre dans ses images ? Seule l'analyse de la phénoménalité en oeuvre de l'image ouvrira des pistes de réponse.

### 1/ Il n'y a pas de beaux nuages dans le ciel

Lorsque l'on regarde une photographie de ciel nuageux d'Alfred Stieglitz, tel *Equivalent* 1927 exposé au Musée d'Orsay, notre appréciation dira que l'on voit un beau ciel, et même — métaphoriquement - un ciel expressif, ne sachant pas pourquoi nous qualifions ce ciel ainsi, si ce n'est peut-être à cause du noir et blanc, renvoyant à une forme expressionniste en art, de rapport ombre/obscurité lié au contraste.

Ce ciel est beau. Mais il ne peut l'être que pour le sujet que je suis face à l'objet de la perception que je vois : le ciel n'est pas beau en soi. Phénoménologiquement, il est beau *pour* un sujet percevant : en tant que phénomène perçu. Le beau, perçu dans la nature, est lié au jugement du sujet, à sa formation, aux structures apprises, analogiquement à un *artefact*. *En soi*, il n'y a pas de beau, comparativement à ce qu'entend l'homme par beau. Ou bien ce serait être dans une tentation anthropomorphique et penser que la nature, comme sujet agissant, aurait façonné les choses selon une téléologie esthétique conforme à la volonté et à la sensibilité humaine[10]. Ionesco l'énonçait : « Un arbre est un arbre, il n'a pas besoin de mon autorisation pour être un arbre; l'arbre ne se pose pas le problème d'être un tel arbre, de se faire connaître comme arbre. Il ne s'explicite pas.

---

[8] Joaquim Sivestre, Yasushi Ikeda et François Guéna, ARTIFICIAL IMAGINATION OF ARCHITECTURE WITH DEEP CONVOLUTIONAL NEURAL NETWORK, Living Systems and Micro-Utopias: Towards Continuous Designing, Proceedings of the 21st International Conference of the Association for Computer-Aided Architectural Design Research in Asia CAADRIA 2016. https://papers.cumincad.org/data/works/att/caadria2016_881.pdf

[9] Grégory Chatonsky, Chatonsky, G. (2020). Finitudes de l'imagination artificielle. Espace, (124), 24–29. « La machine produit des médias de médias à partir de nos données massives qui sont des traces mémorielles du monde. »

[10] « Le spirituel seul est vrai. Le beau naturel est donc un réflexe de l'esprit. Il n'est beau que dans la mesure où il participe de l'esprit. Il doit être conçu comme un mode incomplet de l'esprit » (*ibid.*, p. 11)

Il existe et se manifeste par son existence même »[11]. L'arbre n'est pas beau en soi, mais il est beau pour la conscience percevante.

Mais surtout, le ciel nuageux d'Alfred Stieglitz est beau car ce que je vois n'est pas seulement un ciel nuageux, mais une photographie (un cadre, une profondeur, un contraste, un instant) prise selon une intention humaine. Le beau est lié à l'intention du regard de l'artiste, qui extrait de l'espace et du temps qui lui font face cela que je vois exposé.

Hegel définit l'esthétique comme l'étude du beau lié à l'esprit humain. « Le beau artistique est plus élevé que celui de la nature. (…) La beauté artistique est la beauté engendrée et réengendrée par l'esprit. (…) Du point de vue formel, n'importe quelle mauvaise idée qui passe par la tête d'un homme est néanmoins plus élevée que n'importe quelle production de la nature, car elle possède spiritualité et liberté ».

Hegel à la suite de cette citation explique en quel sens, le soleil, bien qu'il soit nécessaire, en tant que contenu, puisqu'existant, cependant, en elle-même, cette forme n'est ni libre, ni intentionnelle, et en ce sens elle ne peut être belle. Alors que n'importe quelle mauvaise idée, n'importe quel dessin d'un petit enfant, d'un soleil ou d'un nuage, ravira ses parents, du fait que la forme émerge de l'acte intentionnel d'une conscience.

Dès lors se pose de savoir, si une vision de ciel ou de paysage était seulement liée à la machine, alors sans doute ferions-nous face à une forme étrange, perturbante. C'est ce qu'a visé par exemple Micheal Snow avec **La région centrale**, : une caméra robot, qui assujettie à un bras mécanique rotatif, suit des mouvements au hasard. Ce qui ne désoriente pas seulement le spectateur par rapport à la notion de paysage, d'échelle, mais ce qui le désoriente quant à son jugement, à sa position humaine de perception. Ainsi, Chantale Ackerman, répondit lorsqu'on lui dit que ses travellings faisaient penser à ceux de Snow : « Michael Snow construit des dispositifs machiniques, déshumanisés, mécaniques. »[12].

La notion d'esthétique s'est constituée ainsi face aux produits de l'intentionnalité humaine, face aux oeuvres de l'esprit. Dès lors si une représentation est issue d'une machine ou bien d'une IA, se pose la question d'une étrangeté. Le *kinoglaz* automatisé de Michael Snow ouvre à la question de l'*Uncanny Valley* théorisé par Mashiro Mori en 1970[13]. C'est cette étrangeté qu'il nous faut réfléchir, bien entendu dans d'autres termes que que ceux de Mashimo Mori, au sens où l'étrangeté de ce qui est généré provient sans doute de la nature même du généré et non pas des choses ou contenus générés. Ainsi, par exemple si Sigismond de

---

[11] Ionesco, Notes et contre-notes, p.80, Folio essais, 1966.

[12] https://www.critikat.com/dvd-livres/livres/la-region-centrale-de-michael-snow/

[13] Karl F. MacDorman, « La Vallée de l'Étrange de Mori Masahiro », *e-Phaïstos* [En ligne], VII-2 | 2019, mis en ligne le 15 octobre 2019, consulté le 13 février 2024. URL : http://journals.openedition.org/ephaistos/5333 ; DOI : https://doi.org/10.4000/ephaistos.5333

Mallerey dans son article *La uncanny Valley ou les fausses notes de l'IA*[14] explique que les assistants vocaux automatisés par IA traduisent « un manque de sociabilité flagrant de la part de l'IA », c'est que d'abord et avant tout, les IA ne sont pas des êtres qui se sont développés selon une inter-subjectivité et une inter-relationnalité affective. En ce sens, la mention de l'autre, d'autrui est une mimétique vide ontologiquement de toute altérité, mais qui se produit selon une conscience algorithmique par contre des données soumises par l'homme.

### *2/ L'imagination humaine, l'imagination artificielle et la paréidolie*

« On veut toujours que l'imagination soit la faculté de former des images. Or elle est plutôt la faculté de déformer les images fournies par la perception, elle est surtout la faculté de nous libérer des images premières, de changer ses images »[15].

Bachelard insiste sur la puissance de « l'imagination matérielle » non pas comme productrice d'images, mais comme déformation de l'image. L'imagination ne serait pas d'abord reproductrice (donc dépendante des détails matériels), mais combinatoire : elle mettrait en jeu des éléments non reliés, elle redistribuerait le réel pour en saisir une nouvelle image, un fond qui tiendrait du sujet et qui formerait une nouvelle image du monde. C'est pourquoi, elle permettrait de nous libérer de l'image première donnée, des images changeantes du monde[16].

La *paréidolie* est un trait caractéristique du travail de notre imagination, mais elle implique aussi phénoménalement un autre processus. La paréidolie (construit sur para et eidolon) est le processus de la conscience d'un surcroit de vue dans le voir, dans une perception. Il y a une redoublement dans la perception d'une chose informelle, une forme connue, liée au travail spontané et non dirigé d'un processus d'imagination créatrice. Faculté de déformation, certes, mais selon un premier apprentissage de forme de la part de la conscience. Ce n'est pas pour rien que la paréidolie sera utilisée psychologiquement par Rorschach, afin de faire émerger certaines associations inconscientes du sujet dans l'espace conscient. La paréidolie amène à révéler ce qui s'est constitué ou ce qui se trame dans l'espace latent de la conscience humaine. Ce que l'on découvre déjà avec les buvards de Cozens[17], pour qui les tâches d'encres sur un buvard étaient des déclencheurs d'idée afin que l'artiste puisse explorer son imaginaire. La paréidolie est un processus qui nous révèle le coeur du sujet regardant. Pour quelle raison des formes émergent-elles de l'espace latent de la conscience, cela vient de la formation, de la conscience elle-même, de sa propre histoire, de sa mémoire, de ses affects.

---

[14] https://www.vokode.com/uncanny-valley-fausses-notes-ia/ . Etienne Mineur note le 18 février 2024, en commentaire d'une video produite par Sora, la nouvelle IA openAI pour la video : "J'adore cette vidéo générée par Sora (d'OpenAi), cela me procure curieusement un sentiment plutôt plaisant d'uncanny valley, je n'ai pas de sentiment de rejet comme avec les visages. »

[15] Bachelard, *L'air et les songes - Essai sur l'imagination du mouvement.* p.7, Ed. José Corti, 1943. « c'est avant tout la faculté de nous libérer des premières images, des images changeantes ». (p.10)

[16] « on la rêve substantiellement, intimement, en écartant les formes, les formes périssables, les vaines images, le devenir des surfaces ; elle a un poids, elle est un cœur », L'eau et les rêves, p 129, José Corti, 1941.

[17] cf. Ernst Gombrich, Art and Illusion: a study in the psychology of pictorial representation, London, Phaidon, 2002, p. 154-155

De ce fait, si historiquement l'émergence des images générées par IA sont reliées à la paréidolie, de quel processus cela révèle-t-il ?

En mai 2015, avec DeepDream, les IA génératives vont être associées au concept *d'inceptionnisme* par Alexander Mordvintsev et Mike Tyka[18], le concept de paréidolie apparaîtra immédiatement pour définir ce travail, le mois suivant[19].

Deepdream est le premier moment de l'émergence de l'imagination artificielle au niveau des IA génératives de nouvelles générations.

Deepdream est pensé à partir du travail d'ImageNet et de son Large-Scale Visual Recognition Challenge, mais en inversant le processus. Alors que le réseau de neurone d'ImageNet avait pour but de reconnaître des images (perception artificielle), Deepdream, à partir d'un bruit aléatoire, a pour but de produire des images (imagination artificielle). En quelque sorte, on force le réseau de neurone à *halluciner* une forme à partir d'un *bruit* ou bien d'une image qui n'est pas en relation avec ce qui sera généré.

Si comme nous l'avons expliqué la *paréidolie* amène à questionner un espace latent de la conscience, la *paréidolie artificielle* nous amène à réfléchir l'*imagination artificielle* selon une même logique.

Tout d'abord nous retrouvons avec l'émergence de ces images une forme d'étrangeté. Ces images sont artistiques, mais elles paraissent artistiquement étranges. Memo Akten écrit d'emblée dans son article relatant sa découverte : « It's trippy, surreal, abstract, psychedelic, painterly, rich in detail. » En effet, les images données par Alexander Mordvintsev, notamment celle d'un ciel nuageux déroge à la paréidolie humaine, plus précisément la réinvente processuellement. La paréidolie artificielle, lorsque l'on regarde les premières images produites, ne se constituent pas selon l'unité d'une entité émergente (réduction), mais c'est la totalité de l'espace qui est muté ou halluciné (dispersion, diffusion). Alors que dans la paréidolie humaine, il apparaît qu'il y a abstraction du motif (ce qui a été le moteur de la lecture des taches de Rorschach), *Deepdream* vient épuiser la totalité de l'espace de l'image ou du bruit. C'est en ce sens que les images sont des entrelacements de motifs. Deepdream recherche les motifs appris par épuisement à chaque récursion. C'est en ce sens que le réseau de convolution, amène aussi à une forme d'hallucination de l'hallucination. Il n'y a pas d'abstractions de forme, mais dispersion de reconnaissance, travail infini de la récursion tant qu'elle n'est pas arrêtée.

Dès l'émergence de *Deepdream*, se dessine, par les termes employés pour décrire ce processus, qu'une forme d'incompréhension humaine du processus de l'imagination artificielle se constitue. C'est pourquoi beaucoup d'articles auront la tentation de parler d'une forme d'autonomisation, de hiatus entre l'agent humain et un agent artificiel. Avec l'IA, *la boîte noire* de l'appareil technique s'est un peu plus refermée face à l'appréhension de notre conscience stupéfaite.

---

[18] Alexander Mordvintsev, Mike Tyka, Inceptionism: Going Deeper into Neural Networks, juin 2015, https://blog.research.google/2015/06/inceptionism-going-deeper-into-neural.html

[19] cf. Par exemple ce qu'écrit dès le 10 juillet l'artiste Memo AKten : #Deepdream is blowing my mind, https://memoakten.medium.com/deepdream-is-blowing-my-mind-6a2c8669c698

### *3/ Médiation : l'invidence de la boîte noire*

Lorsque nous utilisons des IA génératives, nous ignorons pour la plupart le fonctionnement de celles-ci. Certes avec un colab[20] de Google, nous comprenons davantage la logique de production, puisque nous devons initiés chaque étape : des algorithmes généraux en python, aux modèles, aux génération. Mais ce n'est qu'un aperçu superficiel et trompeur qui donne accès seulement à des étapes et quelques variables mais non pas au coeur de ce qui se produit. Avec Dall-e, mid-journey ou stable diffusion, l'obscurité est renforcée, l'utilisateur reste aveugle face à des processus qui lui sont totalement voilés par l'interface. Tel que l'énonce Benoît Georges en 2017 : « Le problème, c'est que personne, pas même leurs concepteurs, n'est en mesure d'analyser par quel raisonnement ils arrivent à de si bons résultats. Les spécialistes de l'IA appellent cela *« l'effet boîte noire »*. »[21]

Pour saisir cette difficulté de la médiation technique, il semble judicieux de revenir à ce qu'explique Vilem Flusser dans *Pour une philosophie de la photographie* : « les appareils sont des boîtes noires qui stimulent la pensée humaine »[22]. Les appareils techniques se referment face à la compréhension humaine. Dans ce texte Flusser appréhende l'appareil photographique, et il met en tension l'homme et l'appareil technique. Il détache l'appareil de la seule instrumentalité, il l'autonomise du point de vue de fonctionnalité qui sont hétérogènes aux intentionnalité proprement humaines. Ainsi, si « L'appareil fait ce que veut le photographe, (…) le photographe doit vouloir ce que peut l'appareil. (…) Le geste du photographe comme recherche d'un point de vue sur une scène prend place au travers des possibilités offertes par le dispositif. Le photographe se déplace au sein de catégories spécifiques de l'espace et du temps par rapport à la scène : proximité et distance, vues frontales et de côtés, exposition courte ou longue, etc. (…) Ces catégories sont un a priori pour lui. Il doit « décider » à travers elles : il doit appuyer sur le déclencheur. »

Pour Flusser, le photographe n'est pas libre de ce qu'il veut face à la machine, il doit comprendre les catégories impliquées par l'appareil. En quelque sorte il met en évidence une forme de transcendantalité de la machine que doit mettre au jour le photographe dans sa pratique.

Il est certain, qu'avec l'avènement des technologies numériques, la résistance de la boîte noire s'est renforcée, au sens où une double obscurité enveloppe les usages courants des technologies du numérique : obscurité à la fois de la couche programmée et de la couche matérielle. Loin de toute évidence, une in-vidence processuelle apparaît constitutive des imaginations artificielles. De plus, une forme d'*agentivité* numérique s'est constituée, au sens où le dispositif numérique semble

---

[20] Les colab de Google sont des suite de scripts python que l'utilisateur peut modifier, et qu'il va lancer successivement. Cela permet de comprendre les étapes du processus de l'IA, mais en aucun cas ce qu'a été l'apprentissage profond de la machine, les liaisons sémio-éidétiques qui ont été conçus et enfin comment va se constituer l'hallucination de l'image.

[21] Benoit Georges, Les Échos, 15 mai 2017, https://www.lesechos.fr/2017/05/le-talon-dachille-de-lintelligence-artificielle-168099

[22] Vilem Flusser, *Pour une philosophie de la photographie*, ed. Circé, 1996.

agir par lui-même dans un ensemble d'opération qui précédemment n'était relié qu'au photographe.

Laurence Danguy et Julien Schuh on raison alors de souligner que : « L'œil numérique peut ainsi désigner l'ensemble des médiations nouvelles qui créent une culture visuelle devenue hybride, puisque la cognition est partagée et répartie (distribuée) dans des systèmes numériques entre agents cognitifs humains et non-humains »[23]. Reste la distinction d'une double agentivité, exige de saisir *spécifiquement* la nature de l'agent technologique, ce que nous nommons la boîte noire.

C'est dans cet horizon que nous proposons d'introduire la question de la transcendantalité de l'IA. Ici, cependant, nous ne renvoyons pas à une transcendantalité liée à un cogito synthétique apriori, au sens kantien ou bien husserlien. Mais nous nous posons davantage dans la perspective de Gilbert Simondon et de sa redéfinition de la transcendantalité, tel que Deleuze dans *la logique du sens* la met en perspective : « déterminer un champ transcendantal impersonnel et pré-individuel, qui ne ressemble pas aux champs empiriques correspondants et qui ne se confond pas pourtant avec une profondeur indifférenciée. Ce champ ne peut pas être déterminé comme celui d'une conscience... Une conscience n'est rien sans synthèse d'unification, mais il n'y a pas de synthèse d'unification de conscience sans forme du Je ni point de vue du Moi »[24].

Il s'agit de saisir certaines conditions apriori des IA, non pas unifiées sous l'unité d'un noyau de conscience et de réflexivité, mais selon une représentation pré-personnelle et pré-individuelle.

### *4/ Promptologie et rupture du corrélationnisme empirique*

L'ignorance constitutive de l'usage est à la base des recherches qui se constituent à travers les promptologies. La promptologie est le fait d'insérer dans le processus d'imagination artificielle, une forme intentionnelle pour la création : un énoncé. Du fait de l'*invidence* de l'IA, il y a une forme de croyance en un déterminisme entre l'intention du prompteur et le résultat généré. L'image serait causalement le résultat de l'énoncé.

Le premier point qui amène un réductionnisme phénoménologique tient à la prétention d'une maîtrise causale entre le prompt (l'énoncé intentionnellement produit par un humain) et le résultat généré par l'IA. La conscience humaine tend à restreindre le système causal, comme si le résultat était strictement lié à son énoncé. Le développement des conseils en promptologie tendent à poser cela de plus en plus. De même que l'amélioration des modèles de génération d'images et des modèles de compréhension linguistique.

Cependant, il faut envisager que l'IA, en tant qu'entité pré-individuelle, ne se structure pas comme un noyau, mais selon une *logique réticulaire*, un réseau de flux d'informations, de couches et de strates de formations, elle est une structure complexe fondée sur plusieurs étapes et couches constitutives. Nous pouvons synthétiser cela selon deux grandes phases : une phase *éidétique* et une phase *sémiotique*. Le prompteur fait comme si il y avait une naturalisé de lien entre son input (le prompt[25]) et l'output de l'IA. Mais, le lien est de convention et ne repose que sur une analogie. En aucun cas le prompteur peut se dire être le créateur de l'image. Analogiquement, dans une sphère anthropologique relationnelle, c'est comme si en donnant un sujet de dessin ou de photo à une autre conscience, on s'appropriait le résultat. Le prompteur, dans la très grande majorité des cas : n'a ni établi les modèles, ni compris comment se constituait la génération[26].

---

[23] Laurence Danguy, Julien Shuh, L'oeil Numérique : vers une culture visuelle hybride, Sociétés et représentations n°55, éditions de la sorbonne, 2023, p.69.

[24] Deleuze, *Logique du sens*, Minuit, pp.139.

Il y a là une ignorance de la boîte noire flusserienne que nous avons mis en évidence quant au processus d'induction statistique et de liaison sémio-éidétique.

La conscience humaine ici réduit la dimension de l'espace latent à être conforme apriori avec les caractéristiques intentionnelles de la conscience humaine. Ce réductionnisme est lui-même redoublé par un deuxième apriori. 1/ La conscience humaine croit que l'IA est réduite à sa propre intentionnalité. 2/ De plus, elle ne se rend pas compte que les structures intentionnelles et représentationnelles de son intention sont hétérogènes avec l'IA et ses processus d'imagination artificielle.

L'imagination artificielle n'est pas de même nature que l'imagination humaine, il y a une rupture de ce que l'on pourrait constituer à partir de Peirce comme logique indicielle de l'image. Comme le précise Umberto Eco[27], la photographie du point de vue contemporain, mais en quelque sorte toute image produite par l'homme, enveloppe une *indicialité* par rapport au réel. Une image est une trace de réel, et d'autant plus une photographie. Le lien n'est pas seulement par ressemblance ou par analogie.

Le prompt humain, ne définit pas des objets *in abstracto*, tout énoncé enveloppe aussi *l'indicialité* d'un domaine empirique.

Si par exemple, je considère un sac à main[28], intuitivement[29], le sac est un volume, un contenant. Sa *cause finale* est de pouvoir contenir des éléments. C'est d'ailleurs ce que décrit Marie Desplechin dans *Le sac à main*. Le sac est une intériorité où sont déposés et contenus des objets intimes[30]. En ce sens toute représentation d'un sac, renvoie pour le spectateur à cette réalité empirique du sac. Cette *indicialité* de l'image a été originellement soulignée dès Platon, qui dans la *République*, montrait au livre X, le *phantasmata* de l'*eidos* de l'image, en quel sens elle nous trompait, amenant qu'elle n'est pas perçue comme image, mais que l'image s'effaçait au profit de notre regard sur les objets représentés.

---

[25] Il faut noter ici qu'un prompt n'est pas une phrase usuelle. Une phrase usuelle si elle peut être utilisée en tant que prompt, toutefois elle ne réfléchit aucunement les spécificités sémiotiques et syntaxiques permettant un dialogue avec les models.

[26] Pour ma part, c'est ici que mon travail artistique s'est constitué. Depuis 2021 et l'usage des colabs, je constitue systématiquement une partie des modèles que j'utilise. Ce qui a été renforcé par l'usage de Stable-diffusion, qui permet à plusieurs étapes des rendus, d'implémenter des modèles que nous pouvons créer.

[27] Umberto Eco, *Critique of the image, in Thinking Photography, ed. Communications and culture, 1975, p. 32.* « De Peirce, en passant par Morris, aux différentes positions de la sémiotique aujourd'hui, on parle allègrement de ce signe emblématique comme d'un signe possédant certaines propriétés de l'objet représenté. »
.

[28] Éléments de recherche présentés en octobre 2022 à Casablanca lors du colloque Universitaire …

[29] Nous sommes ici face à une dimension anté-prédicative pour le sujet. L'intuition du volume et de la profondeur est constitutive de l'aperception de l'objet, elle fait partie en quelque sorte de la forme apriori de la perception de l'espace.

L'imagination artificielle, lorsqu'on lui demande de représenter un sac, elle n'enveloppe pas un ensemble de catégories liés à notre représentation du sac. S'il y a un bien un corrélationnisme propre à l'IA, celui-ci se constitue en relation avec l'apprentissage profond (deep Learning) des images (modèle général), voire avec les petits modèles d'inférence (Lora : les low rank adaptation). Son expérience est celle de l'apprentissage d'image plates, qui ne sont pas reliées à une expérience de corps. L'espace latent est en ce sens autonome de toute sensibilité du monde au sens humain.

Ici apparaît un des points essentiels pour ces prolégomènes phénoménologiques de l'imagination artificielle : il y a *indicialité* de l'image par rapport à l'expérience pour l'homme car nous sommes existentiellement reliés à un monde empirique par un corps, une sensibilité, une perception. Se découvre une *différence ontologique* de l'expérience d'apprentissages entre IA et conscience humaine, différence par laquelle se constitue le point aveugle de la *promptologie*. Phénoménologiquement, comme l'exprime Husserl, l'être humain est un *Leib* (corps de chair) « à la fois le corps dans sa stature, sa forme spatiale organique et l'intime du rapport au vivre : il est lieu d'inscription du sensible […] – et quasiment le lieu de l'âme »[31]. L'IA n'a pas de *Leib*. Son rapport aux images n'est pas lié à une perception sensible d'objets dans des relations praxiques, mais à l'analyse stochastique de données images et à la constitution par induction statistique de modèles génératifs sémio-éidétiques. Le monde de l'IA est celui des données et jamais celui d'un réel surgissant pour un Leib sensible ouvert. Les structures transcendantales de l'homme doivent être conçues dans cette rencontre avec l'expérience, alors que celles de l'IA sont reliées aux processus réticulaires informatiques de sa constitution.

Dès lors ce qu'il faut souligner c'est que l'IA n'est pas en relation avec le domaine empirique. Le prompt n'est pas un énoncé enveloppant une relation à l'expérience, et de là à des mécanismes intuitifs, antéprédicatifs, de synthèse d'aperception de la sensibilité en un sens husserlien, mais l'IA repose sur un processus d'apprentissages profonds constitués de données analysées pour produire une forme d'induction statistique.

---

[30] Marie Dépléchin, Le sac à main, ed Points, 2006 : « Une jeune femme dresse l'inventaire de son sac à main. Un bâton de rouge à lèvres, un paquet de mouchoirs, un agenda, une liste de courses, un préservatif, une boîte d'allumettes... Chaque objet évoque une histoire, des visages, des voyages, des rêves enfouis ; chacun reflète la vérité intime d'une femme en quête d'elle-même. »

[31] Husserl, *Ideen* II, p.408. Cf. Philippe Boisnard, L'IA et le pararéalisme de la forme (2021) http://databaz.org/xtrm-art/?p=853.

Si Claire Chatelet a raison de préciser citant Benjamin que : « La légende ne deviendra-t-elle pas l'élément essentiel de la prise de vue ? »[32]. En effet, si dans la photographie : « Ici doit intervenir la légende, qui engrène dans la photographie la littéralisation des conditions de vie, et sans laquelle toute construction photographique demeure incertaine. », si le prompt renvoie à une réalité de donnée hétérogène aux processus de modèle des IA, alors toute image de IA, reste incertaine, enveloppe une incertitude quant à ces relations avec un réel intuitivement présent pour l'homme regardant l'image.

C'est pourquoi on ne peut détacher le travail de création d'image de l'IA de ce qu'est sa perception et de la logique de diffusion, à savoir de la paréidolie artificielle reposant sur une analyse probabiliste[33].

En un sens kantien s'il y a une forme de schématisme spécifique permettant la liaison entre le domaine empirique et la conscience qui passe par l'imagination productrice selon Kant, et que celui-ci tel qu'il le dit reste difficile à cerner[34], le *schématisme* à analyser pour l'IA est d'une autre nature, car l'imagination artificielle n'infère pas la forme générale d'un objet d'une expérience à un monde d'objets matériels et pratiques.

### *5/ Réalisme, surréalisme et pararéalisme (pour une ontologie esthétique de l'IA)*

Dès lors qu'il s'agit de penser distinctement l'imagination artificielle, de la considérer comme une *machine célibataire*, il paraît nécessaire de questionner certaines des catégories qui constituent notre discours par rapport aux images qui sont produites par elle.

Le *réalisme pictural* d'une certaine manière a été pour la première fois formulé par Platon. Il est lié à une mimétique esthétique de la représentation face au monde perçu. C'est en ce sens que son analyse s'intéresse au trompe l'oeil de Zeuxis, à ces fameux raisins pouvant même berner un oiseau au point qu'il se casse le bec contre le mur de la représentation. Cette première approche insiste — certes d'un point de vue critique — sur le fait que la représentation graphique est une imitation de ce qui est perçu, un *eidos*, pouvant même concurrencer le réel. Le réalisme en peinture, en Occident s'établit et trouve quant à ses règles de représentation, ses formulations canoniques, avec la Renaissance tout d'abord avec le *De pictura* d'Alberti, établissant les règles de la perspective à partir de la composition du point de fuite et de l'axe de perception de l'observateur (primat de la subjectivité et ontologie du réel), ensuite avec la formulation non achevée de Léonard de Vinci dans son *Traité de La peinture*. « Il faut qu'à force de talent il se donne à lui-même l'ombre, la lumière, la perspective, qu'il se convertisse en la nature même. L'oeuvre représente directement les œuvres de la nature, elle n'a besoin ni d'interprètes, ni de commentateurs. » énonce Léonard de Vinci.

---

[32] Claire Chatelet, Autour de l'intelligence artificielle générative : qui fait l'image, que fait l'image ? Et Xalter Benjamin.

[33] La création des modèles reposent sur des analyses stochastiques, les algorithmes stochastiques sont essentiellement des techniques de simulation de lois de probabilit´es complexes sur des espaces de grandes dimensions. En ce sens l'association sémio-éidétique de l'IA n'est pas liée à une expérience empirique et à une forme d'Einfulung lié à un Leib. C'est ici une des différences fondamentale et essentielles, qu iamènent que la modélisation spécifique sémio-éidétique demande une très grande quantités de données pour que l'analyse stochastique puisse aboutir à un résultat pertinent. Si l'on ne donne qu'une image de chien, seul ce chien là sera reconnu et pourra être conçu.

[34] « Ce schématisme <...> est un art caché dans les profondeurs de l'âme humaine, dont nous arracherons toujours difficilement les vrais mécanismes à la nature pour les mettre à découvert devant nos yeux. » Kant, Critique de la raison pure, p.141 (A141/B180).

Le réalisme de la représentation repose sur l'imitation, et donc sur une ontologie où un réel préexisterait. Le degré de réalisme est lié à la comparaison avec la nature et ses règles. « Le relief donne à l'image l'intensité du réel, par lui seul l'art égale la nature. La peinture est une sorte de magie : Le tableau doit apparaître comme une chose naturelle vue dans un grand miroir. »

Le réalisme tire ses règles de la perception du monde et de la Nature, au sens géométrique. C'est pour cela que le terme d'imitation est si important chez Vinci, même s'il convoque l'imagination. L'imagination est dominée par une ontologie première du monde qui est celle du réel et de ses règles.

Avec l'avènement des technologies de captation, notamment la photographie puis le cinéma, le réalisme s'est peu à peu déplacé dans une correspondance avec la technique. L'exposition *Enfin le cinéma*, au Musée d'Orsay montre parfaitement comment la peinture elle-même a été transformée dans sa manière d'appréhender et d'imiter le réel par la captation technologique – avant même le XXème siècle, par l'introduction du flou ou du plan comme l'indique parfaitement Dominique Païni dans le catalogue.

André Breton avec le *Manifeste du surréalisme*, vient s'opposer et dépasser l'ontologie première du matéralisme et du réalisme. « Cette imagination qui n'admettait pas de bornes, on ne lui permet plus de s'exercer que selon les lois d'une utilité arbitraire; elle est incapable d'assumer longtemps ce rôle inférieur et, aux environs de la vingtième année, préfère, en général, abandonner l'homme à son destin sans lumière. La seule imagination me rend compte de ce qui peut être, et c'est assez pour lever un peu le terrible interdit.
Le procès de l'attitude réaliste demande à être instruit, après le procès de l'attitude matérialiste. Celle-ci, plus poétique, d'ailleurs, que la précédente, implique de la part de l'homme un orgueil, certes, monstrueux, mais non une nouvelle et plus complète déchéance. Il convient d'y voir, avant tout, une heureuse réaction contre quelques tendances dérisoires du spiritualisme. Enfin, elle n'est pas incompatible avec une certaine élévation de pensée."

Breton va défendre le merveilleux dans la littérature et ceci après une analyse du rêve. Le surréalisme se construit par la rupture de la réduction du monde à la logique rationnelle et fonctionnelle. Il va donner un primat à l'imagination comme principe de regard et de constitution de monde. Alors que chez Vinci, le réalisme tirait ses règles de la nature, le surréalisme doit tirer ses règles de l'imagination pour se saisir de ce qui se donne dans la nature. Le surréalisme ne peut s'émanciper de la position première du réalisme. Mais il pose une rupture entre le réel rationnel (au sens de Hegel) et le travail autonome de l'imagination créant ses propres règles.

Grégory Chatonsky dans *Art press* 492, interviewé par Dominique Moulon, parle à propos des créations produites par IA « de réalisme sans réel ». Son expression pertinente, toutefois reste quelque peu problématique.

Le réel, pour le peintre ou le photographe, est ce qui se donne ontologiquement face à lui à travers une manière de percevoir. Quelque soit la figuration, tel que l'énonce Philippe Descola : « figurer, c'est ainsi donner à voir l'ossature ontologique du réel, à laquelle chacun de nous se sera accommodé en fonction des habitudes que notre regard a prises de suivre plutôt tel ou tel pli du monde »[35]. Les surréalistes ne s'échappent pas de ce constat. Voulant dépasser le réel matérialiste et scientiste par la liberté d'un autre réel : issu de l'imagination, il constitue un surréel établi relativement au réel extérieur ontologiquement premier. Dans les deux cas, il y a une ontologie du sujet humain qui est posée, et d'autre part une forme de liaison avec une ontologie générale du réel.

On ne peut penser, lorsque l'on considère une IA, qu'elle produirait *un réalisme sans réel*. On associe deux points de vue dans cette expression : celui de l'IA et celui de l'observateur humain. Il est nécessaire de tenter de réfléchir d'une manière autonome la production de l'IA, en *soustrayant* la question de la perception humaine.

Pour une IA, le réel, n'est plus à entendre comme ce qui fait face au sujet humain répondant d'une ontologie générale, mais comme ce qui constitue ses données : le réel ce sont les données à partir desquelles elle s'entraine, qu'elle va analyser, dont elle va constituer sa mémoire. Penser les réalisations de l'IA sous l'analogie de la photographie est un leurre. Sa perception n'obéit pas à la perception humaine, mais va s'établir dans des calculs statistiques complexes. Ce réel, ne correspond pas à nos règles de perception mais il obéit aux règles algorithmiques qui lui sont propres : les images ne sont pas son réel seulement, mais c'est l'ensemble des données qui en sont issues.

Il y a bien un réel dans l'IA, mais sa définition est hétérogène avec le nôtre. Ses données peuvent être aussi bien des photographies que des graphiques, que n'importe quoi de représentable et qui entre dans la constitution de ses modèles. Pour une IA, la nature de la représentation n'est pas déterminante, ni non plus ce qu'elle produira.

C'est à partir de ce réel que l'IA crée, et non pas à partir de rien. Un réalisme sans réel, cette expression vaut pour l'observateur humain. Cependant ce que produit l'IA, si pour une part cela ressemble à notre réalité, ce n'est aucunement un réalisme, ni un surréalisme. Ce n'est pas un réalisme, au sens où l'IA n'a pas pour modèle une antériorité ontologique de réel, de monde, qu'elle tenterait d'imiter. Elle ne tente pas non plus de s'affranchir d'un réel, en définissant un surréel. Elle explore par sa complexion une puissance graphique (un bruit, quelque soit l'image initiale : un noise, une diffusion, une image concrète) et travaille à créer selon des possibilités statistiquement déterminées par son apprentissage.

L'IA à chaque fois qu'elle produit une image, crée un nouveau réalisme. C'est ce que j'appelle un *pararéalisme*. Un réalisme d'â-côté, comme si l'IA composait une dimension parallèle. L'IA figure sans relation avec un objet à figurer[36].

Le *pararéalisme* est la constitution d'un réalisme sans relation avec la fondation ontologique qui définit le réalisme chez l'homme. Penser que les créations d'une IA seraient surréalistes, reviendrait à poser l'intentionnalité humaine comme motrice, et neutraliser la question de l'IA, en rabattant et réduisant cette dernière au simple instrument. Or, dans les représentations que l'on peut produire avec une IA, quelque chose échappe inexorablement de la volonté du sujet humain : il ne peut être que dans l'expectative de ce qui arrive, dans des esquisses de compréhension, des tentatives d'interaction parcellaire. Cela provient du fait que le réel duquel part l'IA, le sujet humain ne le connaît pas et ne peut même en avoir une idée claire et distincte : puisque ce réel n'est pas tant l'image que le résultat de l'analyse des images pour l'IA, le réseau des interconnections statistiques des images et des textes.

---

[35] Philippe Descola, *Les formes du visible*, ed Seuil, 2021.

[36] Il faudrait alors repenser les quatre catégories établies par Philippe Descola quant à la figuration : analogisme, naturalisme, totémisme, animisme.

### *6/ Quelques aspects d'une approche phénoménologique*

Il s'agit d'élaborer une approche phénoménale de l'imagination artificielle, en tant qu'elle constitue un *pararéalisme* pour la conscience humaine. En ce sens de décrire strictement certains processus qui nous paraissent impliqués par la génération en tant que phénomène afin de déduire certaines règles - sans doute temporaires[37] - de l'image générée. Il semble que le regard que nous avons de l'objet généré enveloppe un ensemble d'apriori, qui ne permet pas spécifiquement de réfléchir une forme de *transcendantalité algorithmique* de l'IA. Cette approche, pour l'instant a été peu explorée.

Notre perception des images est reliée aux formes apriori de notre perception (l'espace et le temps) de même qu'à notre culture. Nous faisons *comme si* ces images étaient faites pour nous et produites par une conscience similaire à la nôtre.

Pour finir cet article, je vais mettre en évidence à travers la question de l'espace et du temps quelques traits spécifiques de l'image de l'IA appréhendée phénoménalement, et indiquer à partir de certaines de mes oeuvres les enjeux qui s'y jouent dans la pratique artistique.

*Ponction - génération - espace*

Dans les millions d'images générés par IA, s'impose le photo-réalisme, une volonté de copier le réel, ou encore la représentation que nous nous faisons du réel[38]. Toutefois tel que le précise, François Cholet, « La génération d'images est une forme de photographie, dans un espace latent qui interpole entre des centaines de millions d'images. Lorsque vous prenez une photo, vous ne la "créez" pas, vous la prenez. Vous trouvez la scène que vous voulez, et vous la capturez comme vous le voulez. Il s'agit d'une forme de sélection. »[39].

---

[37] il serait impossible ici de prétendre à établir des règles définitives, au sens où nous faisons face à l'émergence et la constitution d'une générativité technologique. Les transformations de la compréhension des IA et des modèles et des algorithmes, ont amené un certain nombre de transformations dans les rendus. Toutefois, certains principes nous semblent persistants.

[38] C'est en ce sens ue dans les prompts seront utilisés des termes techniques, liés à la photographie ou au cinéma : HDR, photorealistic, 4K, 8K, Blender, 3D …

[39] François Chollet, dans son compte twitter, 2 septembre 2022 : « Image generation is a form of photography. Photography in a latent space that interpolates between hundreds of millions of images. When you take a photo, you don't "create" the picture, you take it. You find the scene you want, and you capture it the way you want. It's curation. » cité et traduit par Laurence Donguy et Julien Shut « L'oeil numérique : vers une culture visuelle hybride ».

Pour bien saisir la spécificité de l'espace de l'imagination artificielle, nous allons éclaircir la différence entre la forme intuitive de la perception sensible de l'espace chez l'homme, de celle de l'imagination artificielle. Non pour les comparer ou les hiérarchiser, mais pour les différencier spécifiquement.

L'image que le sujet humain capture est une sélection. Même un tableau peint est une partie d'un monde possible, aussi bien pour le peintre que le spectateur. La pensée humaine a l'intuition de l'expansion de l'espace, de sa continuité. Le cadre n'est pas la fermeture d'un monde, mais ouvre à la question du hors-champ, du monde qui enveloppe la partie[40]. Dans toute représentation du réel, peinte, ou photographiée, il y a une forme intuitive de l'espace et de l'étendue. Dès lors quelque soit la mouvance photographique, « reproductrice » ou « formatrice » pour reprendre les termes de Siegfried Kracauer[41], une même condition transcendantale est à l'oeuvre.

Un hiatus se dessine avec l'émergence de l'espace par l'IA. L'image générée n'a pas de hors champ. Elle n'est jamais une ponction dans un monde qui existe. Ce qui est généré est le monde en tant que monde replié sur sa propre logique générative. Rien ne déborde l'image. L'étendue qui est générée n'enveloppe pas l'intuition d'un monde environnant, une ontologie générale du monde. Chaque génération est pour elle-même son propre monde. Cela ne se voit pas à travers la diversité des images qui sont diffusées. Car le spectateur appréhende bien évidemment l'image selon ses propres intuitions de l'espace et du temps. Toutefois, si nous considérons l'image en tant que telle, détachée de notre perception, mais pensée en son émergence processuelle : elle est à elle seule un monde limité à la dimensionnalité première du bruit qui a permis sa génération. Comprendre cela, implique que toute extension de l'image par l'IA n'est pas une extension, mais une juxtaposition générative. L'image est ainsi insulaire. Si une photographie prise par l'homme est toujours une parcelle d'un monde réel plus large, de même qu'une peinture implique la possibilité d'un monde, l'imagination artificielle crée des juxtapositions de monde.

Avec ma série Dreamers, j'ai tenté de mettre en avant cela. L'espace généré, par succession de déplacement des cadres de génération est un assemblage de perspective, de situations. Chaque humain est représenté avec un casque de VR, afin de signifier cette archipelisation du sujet dans son cadre de génération. Pour cette génération, j'ai à chaque fois indiqué qu'il fallait qu ecela soit photo-réaliste. Et pourtant, peu à peu, par la multiplication des points d'horizon, par l'hallucination des liaisons entre les juxtapositions, un pararéel s'est constitué, qui n'est propre qu'à l'IA.

*De la profondeur*

Il y a une rupture du corrélationnisme ontologique de l'IA au réel empirique que nous expérimentons. L'IA n'a pas de forme intuitive de l'espace et du temps relié à un corps qui explore un monde. Ce qu'elle enveloppe est l'analyse d'images, de données. Les concepts (sémiotiques) qui sont articulés (sémantiques) et reliés avec les images (eidétiques) ne sont que des signes arbitraires pour des formes analysées. En ce sens, ils ne désignent pas des expériences sensibles mais des descriptions d'images.

---

[40] cf. Anne Cauquelin, L'invention du paysage, PUF, 1989. « Par la fenêtre, je vois donc quelque chose de la nature, prélevé sur la nature, découpé dans son domaine. Le paysage n'est autre chose que la présentation culturellement instituée de cette nature qui m'enveloppe. » (p. 127)

[41] Siegfried Kracauer, Théorie du film : La rédemption de la réalité matérielle, Flammarion, Paris, 2010.

Il n'y a pas de profondeur pour les IA, et il n'y a pas d'intérieur non plus. Plusieurs de mes créations ont tenté d'approcher cette tension entre d'un côté l'intuition humaine de la profondeur et de l'intériorité et de l'autre, la création de formes sans épaisseur de l'IA.

La série *En quête de*, représente sur chaque photographie générée, des hommes qui font face à des portes, des ouvertures dans la jungle. J'ai repris le thème de la quête de soi, issue de Conrad et de sa relecture dans *Apocalypse Now*. Toutefois, j'ai travaillé à cette suspension face à une profondeur de l'image absente de la génération, mais seulement présente pour l'homme. Il y a ici un basculement entre l'espace latent et sa virtualité et l'espace latent de l'imaginaire humain. Si tout d'abord, il y a délégation par le créateur et son prompt de la figuration à la machine, ensuite, dans le regard, du fait de l'impossibilité de l'intuition de profondeur dans l'image et le processus de génération, il y a dialectiquement une appropriation du contenu possible par le spectateur de l'image.

*La suspension du temps*

Dans la série *The hapenning* que j'ai réalisée en 2022, chaque génération photo-réaliste montre une main tenant un smartphone face à une scène immergée dans de la fumée. De même que l'image n'est pas une ponction dans un espace plus large, la photographie n'est pas une ponction dans le temps.

*The happening*, l'événement, ici est vide. Il n'a aucune épaisseur dans une chaîne causale. Certes le spectateur de la photographie, peut imaginer un avant et un après, mais ils ne sont aucunement viables. Pour l'imagination artificielle : ils ne sont pas transcendantalement établis.

Le temps suspendu de la génération est celui d'un interstice dans les bifurcations du possible de l'espace latent. Il n'y a ni avant, ni après, car toute nouvelle génération enveloppe le possible et non pas le consécutif. Sur ce point, se caractérise la différence par exemple avec la génération de systèmes de particules ou de masses physiques en art numérique. En effet, si on considère par exemple le travail de CHDH dans une oeuvre comme egregore[42], le mouvement des particules est généré à partir d'une équation causale de mouvements obéissant aux équations de l'accélération, vitesse, déplacement de de N*ewton*. On perçoit parfaitement la distinction avec les mouvements créés par les IA génératives, qui sont des interpolations d'images à partir de deep Learning de séquences. Peut-être serait-ce une piste de leur faire induire de l'analyse séquentielle et consécutive d'images des lois de mouvements et de construction.

C'est là un des enjeux majeurs de la recherche en IA générative : la question du temps comme liaison causale. Si pour la conscience humaine, le temps est imbriqué dans la représentation du présent selon un *schématisme* causal, pour l'IA la causalité est constituée de l'analyse d'images consécutives et de l'apprentissage de la consécution des formes, donc de la variation de la forme dans le temps. Début 2024, c'est sur ce point qu'open AI vient encore de franchir une étape avec SORA et l'analyse des consécutions d'images et sa schématisation linguistique reliée à chatGPT[43].

---

[42] https://chdh.net/egregore.php

Cette suspension du temps était éminemment déterminante au tout début des développements des colabs (2021) et des possibilités de création vidéo. Les variations videos permises (zoom, translate XYZ, rotate XYZ) n'étaient ni des exploration d'espace, ni des mutations dans le temps, mais des évolutions par convolution du premier rendu de pixels. C'est pourquoi il y avait une impression de fractals dans ces évolutions videos.

### *Conclusion*

Peut-être jamais dans l'histoire de l'humanité, la conscience humaine n'a-t-elle fait aussi rapidement l'expérience d'une altérité à laquelle elle attribue autant de qualités similaires à elle-même. Toutefois, comme nous venons de le voir, derrière la confusion entraînée par l'invidence du fonctionnement des IA, il paraît nécessaire dès à présent de poser une analyse phénoménologique de ce qui caractérise ce que les IA génèrent. Ceci non pas seulement pour mettre en critique la croyance mimétique de l'IA par rapport à l'homme, mais pour saisir un ensemble de différences entre l'homme et l'IA, notamment s'est apparu quant à la question de l'espace et du temps des représentations.

Nous n'avons établi ici que des prolégomènes à ce qui se présentera alors comme une post-esthétique.

---

[43] cf. Philippe Boisnard, Le temps de l'imagination artificielle, http://databaz.org/xtrm-art/?p=910